\documentclass[aps,pra,amsmath,two column,amssymb,showpacs]{revtex4}
\usepackage{txfonts}
\usepackage{mathbbold}
\usepackage{epsfig,bm,dcolumn}
\usepackage{graphicx}
\usepackage{color}

\begin{document}

\title{Interaction energy and itinerant ferromagnetism in a strongly interacting Fermi gas \\ in the absence of molecule formation}

\author{Lianyi He}

\affiliation{Theoretical Division, Los Alamos National Laboratory, Los Alamos, New Mexico 87545, USA}

\date{\today}

\begin{abstract}
We investigate the interaction energy and the possibility of itinerant ferromagnetism in a strongly interacting Fermi gas at zero temperature in the absence of molecule formation. The interaction energy is obtained by summing the perturbative contributions of Galitskii-Feynman type to all orders in the gas parameter. It can be expressed by a simple phase space integral of an in-medium scattering phase shift. In both three and two dimensions (3D and 2D), the interaction energy shows a maximum before reaching the resonance from the Bose-Einstein condensate side, which provides a possible explanation of the experimental measurements of the interaction energy. This phenomenon can be theoretically explained by the qualitative change of the nature of the binary interaction in the medium. The appearance of an energy maximum has significant effects on the itinerant ferromagnetism. In 3D, the ferromagnetic transition is reentrant and itinerant ferromagnetism exists in a narrow window around the energy maximum. In 2D, the present theoretical approach suggests that itinerant ferromagnetism does not exist, which reflects the fact that the energy maximum becomes much lower than the energy of the fully polarized state.
\end{abstract}


\maketitle

\section {Introduction}\label{s1}

A repulsively interacting Fermi gas can be realized by rapidly quenching the atoms to a metastable state in the absence of bound-state (molecule) formation at the Bose-Einstein condensate (BEC) side of a Feshbach resonance \cite{exp,exp2,exp3,2Dexp}. An important goal is to study the itinerant ferromagnetism in repulsive Fermi systems \cite{IFM01,IFM02,FMTH,FMMC}, which is a long-standing problem in many-body physics. The interaction energy has been measured by studying the expansion properties \cite{exp2} or by using rf spectroscopy \cite{exp3,2Dexp}. In an expansion experiment on a $^6$Li Fermi gas \cite{exp2} around its broad Feshbach resonance at magnetic field $B\simeq834$ G and at temperature
$T\simeq0.6T_{\rm F}$, where $T_{\rm F}$ is the Fermi temperature, it was found that the interaction energy of the repulsive branch suddenly jumps to negative values at magnetic field $B\simeq720$ G, which lies at the BEC side of the resonance. The same feature was also indicated by rf spectroscopy measurement in a two-dimensional Fermi gas \cite{2Dexp,2Dgas}.

A repulsive Fermi gas was previously suggested to exist in the upper branch of a strongly interacting Fermi gas \cite{box}. However, the 
``upper branch" is well defined only for two-body systems. Exact solution of the energy levels of three attractively interacting fermions in a harmonic trap \cite{Hu} shows that there are many avoided crossings between the lowest two branches as one approaches the resonance, making it difficult to identify a repulsive Fermi system. So far there is no precise formulation of it for many-body systems. In this paper, we study a metastable many-body state of a strongly interacting Fermi gas in the absence of molecule formation, or containing only scattering states 
\cite{Ho}. In the high-temperature limit it can be formulated by using the virial expansion to the second order in the fugacity because the 
two-body contribution dominates \cite{virial}. Moreover, in the weak-coupling limit (both the BCS and BEC limits), the equation of state of 
such a system can be described perturbatively \cite{LHY,Galitskii}.

The sudden jump of the interaction energy at the BEC side of the resonance can be qualitatively explained by the strong atom loss around $B=720$ G, where the system may be regarded as a mixture of atoms and weakly bound molecules \cite{exp2}. Shenoy and Ho \cite{Ho} rather suggested that the interaction energy of the repulsive branch was found to increase and then decrease as one approaches the resonance from the BEC side, showing a maximum before reaching the resonance. By using a generalized Nozi$\grave{\rm e}$res-Schmitt-Rink (NSR) approach where the molecular contribution is subtracted, they found that an energy maximum already appears at high temperature $T\sim 3T_{\rm F}$ \cite{Ho}. However, the NSR approach to the repulsive branch is limited to the high-temperature region where the chemical potential becomes negative and the fugacity is small. It becomes less accurate and predicts artificial discontinuities and instability at low temperature \cite{Ho}. Moreover, at low temperature, since the compressibility becomes negative in a large forbidden area, the number equation has no solution and hence the generalized NSR approach cannot provide quantitative predictions.

In this paper, we follow Shenoy and Ho's explanation of the behavior of the interaction energy but employ an alternative nonperturbative approach at zero temperature to overcome the difficulty of a negative compressibility. The basic idea is to sum some certain type of perturbative contributions to all orders in the gas parameter \cite{LRS00,LRS01,LRS02}. The interaction energy ${\cal E}_{\rm int}$ can be formally expressed as
\begin{equation}
{\cal E}_{\rm int}(g)=\sum_{n=1}^\infty c_ng^n,
\end{equation}
where $g$ is the gas parameter. Obviously, the result becomes perturbative at weak coupling $|g|\ll1$. The basic requirement for this resummation is that the interaction energy converges in the strong-coupling limit $g\rightarrow\infty$. According to Bishop \cite{Bishop}, there exist two different schemes to calculate the perturbative equation of state \cite{LHY,Galitskii}, the Bethe-Goldstone scheme and the Galitskii-Feynman (GF) scheme.
If the perturbative contributions can be computed and summed precisely to all orders in $g$, they should agree with each other. However, this is impossible since the problem is not exactly soluble. In this work we employ the GF scheme. In this scheme, both the particle and hole parts of the single-particle propagator are used. By summing the perturbative contributions of the GF type, the contributions from the particle-particle ladders, hole-hole ladders, and mixed particle-particle and hole-hole ladders are resummed self-consistently to all orders in $g$ \cite{LRS02}.

The paper is organized as follows. In Sec. \ref{s2}, we briefly introduce the description of the two-body scattering by using a contact interaction.
In Sec. \ref{s3}, we study the binary scattering at finite density, i.e., in the presence of Fermi surfaces. The interaction energy is calculated in Sec. \ref{s4}. We apply the theory to study the itinerant ferromagnetism in Sec. \ref{s5}. We summarize in Sec. \ref{s6}.

\section {Basics: two-body scattering}\label{s2}

Two-component atomic Fermi gases across a broad $s$-wave Feshbach resonance can be described by the Hamiltonian
\begin{eqnarray}
H=\sum_{\sigma=\uparrow,\downarrow}\int d{\bf r}\psi_\sigma^\dagger({\bf r})\left(-\frac{\hbar^2\nabla^2}{2M}\right)
\psi_\sigma^{\phantom{\dag}}({\bf r})+H_{\rm int}
\end{eqnarray}
with a contact interaction \cite{review}
\begin{eqnarray}
H_{\rm int}=U\int d{\bf r}\psi_\uparrow^\dagger({\bf r})\psi_\downarrow^\dagger({\bf r})\psi_\downarrow^{\phantom{\dag}}({\bf r})\psi_\uparrow^{\phantom{\dag}}({\bf r}).
\end{eqnarray}
Here $\psi_\sigma$ are the fermion fields with $\sigma=\uparrow,\downarrow$ denoting the two components, $M$ is the fermion mass, and $U$ is a contact coupling which represents the short-ranged attractive interaction. The free fermion propagator in vacuum is given by
\begin{eqnarray}
{\cal G}_0(p_0,{\bf p})=\frac{1}{p_0-\varepsilon_{\bf p}+i\epsilon},
\end{eqnarray}
where $p_0$ and ${\bf p}$ denote the energy and momentum of a fermion, $\epsilon=0^+$, and $\varepsilon_{\bf p}={\bf p}^2/(2M)$. For convenience, we use the units $\hbar=M=1$ throughout.

The advantage of using the contact interaction is that the Lippmann-Schwinger equation for the two-body $s$-wave scattering $T$ matrix becomes a simple algebraic equation. In the diagrammatic representation, it is equivalent to resummation of particle-particle ladder diagrams to all orders in $U$. The off-shell $T$ matrix can be expressed as
\begin{eqnarray}
T_{2{\rm B}}(P_0,{\bf P})=\frac{U}{1-U\Pi_0(P_0,{\bf P})},
\end{eqnarray}
where $P_0$ and ${\bf P}$ are the total energy and momentum of the two scattering fermions. The two-body bubble diagram $\Pi_0(P_0,{\bf P})$
is given by
\begin{eqnarray}
\Pi_0(P_0,{\bf P})&=&i\int_{-\infty}^\infty\frac{dq_0}{2\pi}\sum_{\bf q}{\cal G}_0\left(q_+,{\bf q}_+\right){\cal G}_0\left(q_-,{\bf q}_-\right)\nonumber\\
&=&\sum_{\bf q}\frac{1}{P_0+i\epsilon-\frac{{\bf P}^2}{4}-2\varepsilon_{\bf q}}.
\end{eqnarray}
Here we have defined the notations $q_\pm=P_0/2\pm q_0$ and ${\bf q}_\pm={\bf P}/2\pm{\bf q}$. We notice that the two-body bubble function $\Pi(P_0,{\bf P})$ and hence $T_{2\rm B}(P_0,{\bf P})$ depend only on the combination $P_0-{\bf P}^2/4$ because of the Galilean invariance of the two-body system. The scattering amplitude $f(k)$ can be obtained by imposing the on-shell condition $P_0={\bf P}^2/4+E$, where $E=k^2$
is the scattering energy in the center-of-mass frame.

The cost of using the contact interaction is that the integral over ${\bf q}$ becomes divergent. This divergence can be removed through the renormalization of the contact coupling $U$ in terms of the physical scattering length. To this end, we first regularize the divergence by introducing a cutoff $\Lambda$ for $|{\bf q}|$. We obtain
\begin{eqnarray}
\Pi_0(P_0,{\bf P})=-\frac{1}{2\pi^2}\Lambda+\frac{1}{4\pi}\sqrt{-P_0-i\epsilon+\frac{{\bf P}^2}{4}}
\end{eqnarray}
for three dimensions (3D) and
\begin{eqnarray}
\Pi_0(P_0,{\bf P})=-\frac{1}{2\pi}\ln\Lambda+\frac{1}{4\pi}\ln\left(-P_0-i\epsilon+\frac{{\bf P}^2}{4}\right)
\end{eqnarray}
for two dimensions (2D). To renormalize the contact interaction, we match the $T$ matrix $T_{2{\rm B}}(P_0,{\bf P})$ on the scattering mass shell
$P_0={\bf P}^2/4+{\bf k}^2$  to the known scattering amplitude $f(k)$ \cite{review}. In general, we find that only the coupling constant $U$ needs renormalization. In 3D, we have
\begin{eqnarray}
f(k)=\frac{4\pi}{a^{-1}+ik},
\end{eqnarray}
where $a$ is the 3D scattering length. A bound state with binding energy $\varepsilon_{\rm B}=1/a^2$ exists only for $a>0$. The coupling constant is given by
\begin{eqnarray}
U(\Lambda)=-\frac{4\pi}{2\Lambda/\pi-a^{-1}}.
\end{eqnarray}
In 2D, a two-body bound state exists for arbitrarily weak attraction. The scattering amplitude reads \cite{BCS2D}
\begin{eqnarray}
f(k)=\frac{4\pi}{-\ln(E/\varepsilon_{\rm B})+i\pi},
\end{eqnarray}
where $\varepsilon_{\rm B}$ is the binding energy of the bound state. For convenience, we define a 2D scattering length $a_2$. There exist two popular definitions of $a_2$ in the literature. In this paper, we employ the definition $\varepsilon_{\rm B}=1/a_2^2$ in accordance with early theoretical studies \cite{PEOS2D1,PEOS2D2} and recent experimental studies \cite{2Dexp,2Dgas}. Notice that $a_2$ is always positive. From this definition of $a_2$, the coupling constant is given by
\begin{eqnarray}
U(\Lambda)=-\frac{2\pi}{\ln{(\Lambda a_2)}}.
\end{eqnarray}
Another popular definition of the 2D scattering length is given by $\varepsilon_{\rm B}=4/(a_2^2e^{2\gamma})$, where $\gamma\simeq0.577$ is Euler's constant. Converting the theoretical results from one definition to the other is rather simple.

\section {Binary Scattering In medium}\label{s3}

At finite density, the propagators of noninteracting fermions are given by
\begin{eqnarray}
{\cal G}_\sigma(p_0,{\bf p})=\frac{1-n_\sigma({\bf p})}{p_0-\varepsilon_{\bf p}+i\epsilon}
+\frac{n_\sigma({\bf p})}{p_0-\varepsilon_{\bf p}-i\epsilon},
\end{eqnarray}
where $n_\sigma({\bf p})\equiv\Theta(k_{\text F}^\sigma-|{\bf p}|)$. Here $k_{\text F}^{\uparrow,\downarrow}$ are the Fermi momenta of the two spin components. For convenience, we express them as $k_{\text F}^\sigma=k_{\rm F}\eta_\sigma$, where the average Fermi momentum $k_{\rm F}$ is defined by the total density $n=n_\uparrow+n_\downarrow$ and the dimensionless quantities $\eta_\sigma$ depend on the polarization $x=(n_\uparrow-n_\downarrow)/(n_\uparrow+n_\downarrow)$. In 3D we have $n=k_{\rm F}^3/(3\pi)$ and $\eta_{\uparrow,\downarrow}=(1\pm x)^{1/3}$. In 2D, $n=k_{\rm F}^2/(2\pi)$ and $\eta_{\uparrow,\downarrow}=(1\pm x)^{1/2}$. The gas parameter is defined as $g=k_{\rm F}a$ in 3D and
$g=-1/\ln(k_{\rm F}a_2)$ in 2D. It is convenient to use an alternative form of the propagator. It is given by the vacuum-medium decomposition
\begin{eqnarray}
{\cal G}_\sigma(p_0,{\bf p})={\cal G}_0(p_0,{\bf p})+{\cal G}_{\rm m}^\sigma(p_0,{\bf p}),
\end{eqnarray}
where
\begin{eqnarray}
{\cal G}_{\rm m}^\sigma(p_0,{\bf p})=2\pi i\delta(p_0-\varepsilon_{\bf p})n_\sigma({\bf p})
\end{eqnarray}
is called a ``medium insertion" (MI) \cite{LRS02}.

To sum certain types of perturbative contributions, we employ the GF scheme \cite{fetterbook,Galitskii,Bishop}, which takes into account the propagations of both particles and holes and is exact to order $O(g^2)$. The many-body $T$ matrix is given by summation of the GF ladder diagrams 
to all orders in $U$. We have
\begin{equation}
T_{\rm m}(P_0,{\bf P})=\frac{U}{1-U\Pi(P_0,{\bf P})},
\end{equation}
where the bubble diagram $\Pi(P_0,{\bf P})$ is now given by
\begin{eqnarray}
\Pi(P_0,{\bf P})=i\int_{-\infty}^\infty\frac{dq_0}{2\pi}\sum_{\bf q}{\cal G}_\uparrow\left(q_+,{\bf q}_+\right){\cal G}_\downarrow\left(q_-,{\bf q}_-\right).
\end{eqnarray}
According to the vacuum-medium decomposition, it can be decomposed into three parts,
\begin{equation}
\Pi(P_0,{\bf P})=\Pi_0(P_0,{\bf P})+\Pi_1(P_0,{\bf P})+\Pi_2(P_0,{\bf P}),
\end{equation}
where $\Pi_l$ ($l=0,1,2$) stands for the bubble diagram with $l$ MIs. The vacuum contribution $\Pi_0(P_0,{\bf P})$ naturally cancels the cutoff dependence of $U$. The medium contributions are finite and can be evaluated as
\begin{eqnarray}
\Pi_1(P_0,{\bf P})=-\sum_{\bf q}\frac{n_\uparrow({\bf q}_+)+n_\downarrow({\bf q}_-)}
{P_0+i\epsilon-\frac{{\bf P}^2}{4}-2\varepsilon_{\bf q}}
\end{eqnarray}
and
\begin{eqnarray}
\Pi_2(P_0,{\bf P})=-2\pi i\sum_{\bf q}n_\uparrow({\bf q}_+)n_\downarrow({\bf q}_-)\delta\left(P_0-\frac{{\bf P}^2}{4}-2\varepsilon_{\bf q}\right).
\end{eqnarray}
We notice that in the presence of the medium, the two-body bubble function $\Pi(P_0,{\bf P})$ and hence the $T$ matrix depend not only
on the combination $P_0-{\bf P}^2/4$ but also on the momentum ${\bf P}$ through the distribution functions $n_\uparrow({\bf q}_+)$ and
$n_\downarrow({\bf q}_-)$.

The $T$ matrix $T_{\rm m}(P_0,{\bf P})$ characterizes the energy spectrum of the system in the GF approach. We note that the imaginary part of $\Pi(P_0,{\bf P})$ vanishes for $P_0-{\bf P}^2/4<0$. The bound states or molecule states correspond to the poles of the $T$ matrix in the region
$P_0-{\bf P}^2/4<0$. Since we consider only the scattering part of the many-body energy spectrum, which corresponds to the two-particle continuum
$P_0-{\bf P}^2/4>0$, we impose the on-shell condition $P_0={\bf P}^2/4+{\bf k}^2$. For convenience, we define two dimensionless variables
\begin{equation}
s=\frac{|{\bf P}|}{2k_{\rm F}},\ \ \ \ \ t=\frac{|{\bf k}|}{k_{\rm F}}.
\end{equation}
In analogy to the vacuum case, the in-medium scattering amplitude can be expressed as
\begin{equation}
f_{\rm m}(s,t)=\frac{4\pi}{f_1(s,t)+if_2(s,t)}
\end{equation}
where $f_1(s,t)$ and $f_2(s,t)$ are the real and imaginary parts of the denominator, respectively. They can be expressed as
\begin{eqnarray}
f_1(s,t)&=&4\pi\left[U^{-1}-{\rm Re}\Pi\left(P_0=\frac{{\bf P}^2}{4}+{\bf k^2},{\bf P}\right)\right],\nonumber\\
f_2(s,t)&=&-4\pi {\rm Im}\Pi\left(P_0=\frac{{\bf P}^2}{4}+{\bf k^2},{\bf P}\right).
\end{eqnarray}
We note that $f_{\rm m}(s,t)$ recovers the two-body scattering amplitude at vanishing density ($k_{\rm F}^\sigma\rightarrow0$).

In Sec. \ref{s3} we will show that the interaction energy of the many-body scattering state can be expressed in terms of the in-medium scattering phase shift
\begin{eqnarray}
{\cal E}_{\rm int}=-4\pi\sum_{\bf P}\sum_{\bf k}n_\uparrow({\bf k}_+)n_\downarrow({\bf k}_-)
\frac{\phi_{\rm m}(s,t)}{f_2(s,t)},
\end{eqnarray}
where the in-medium scattering phase shift is defined as
\begin{equation}
e^{-2i\phi_{\rm m}(s,t)}=\frac{f_1(s,t)+if_2(s,t)}{f_1(s,t)-if_2(s,t)}.\label{shift}
\end{equation}
At vanishing density, it recovers the two-body scattering phase shift $\phi_{2\rm B}(k)$, where $\phi_{2\rm B}>0$ and $\phi_{2\rm B}<0$ correspond to attraction and repulsion, respectively. In 3D, we have $\phi_{2\rm B}(k)=-\arctan(ka)$. Note that it is different from the usual definition of the scattering phase shift $\phi_{\rm m}=-{\rm Im}\ln(-f_1-if_2)$. From this definition, we have $\phi_{\rm 2B}(0)=\pi$ for $a>0$ and $\phi_{\rm m}(s,t)\rightarrow\pi$ in the BEC limit, which clearly shows the existence of a bound state. Since we are considering a system containing only scattering states, we should exclude the influence of the molecule bound state and use the definition (\ref{shift}).

In the following we analyze the behavior of the phase shift $\phi_{\rm m}$ in the phase space ${\cal S}$ defined as
 \begin{equation}
{\cal S}=\left\{\ ({\bf k},{\bf P})\ \big|\ |{\bf P}/2+{\bf k}|<k_{\rm F}^\uparrow, \ \ |{\bf P}/2-{\bf k}|<k_{\rm F}^\downarrow\right\}.
 \end{equation}
It is crucial for us to understand the behavior of the interaction energy across a Feshbach resonance.

\begin{figure}[!htb]
\begin{center}
\includegraphics[width=9.3cm]{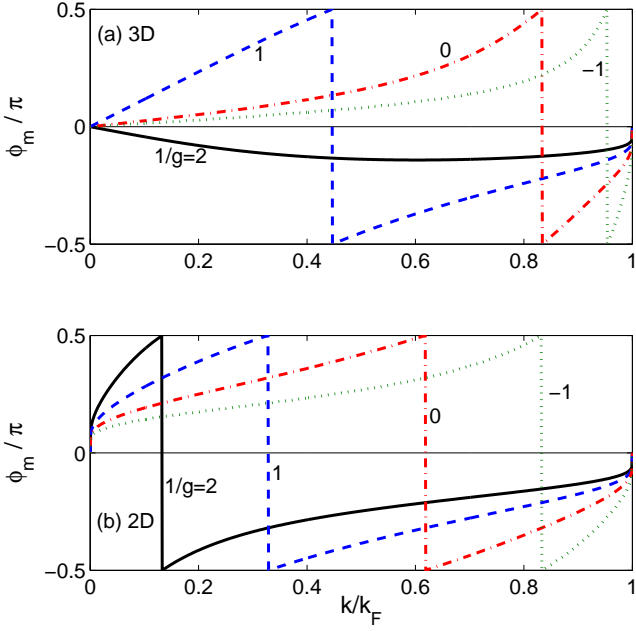}
\caption{(Color online) The in-medium scattering phase shift $\phi_{\rm m}$ at zero center-of-mass momentum ${\bf P}$ for various values of
the inverse gas parameters $g$ in 3D ($g=k_{\rm F}a$)  and 2D [$g=-1/\ln(k_{\rm F}a_2)$]. \label{fig1}}
\end{center}
\end{figure}

\subsection {Three dimensions}

In 3D and in the phase space ${\cal S}$, the functions $f_1(s,t)$ and $f_2(s,t)$ can be expressed as
\begin{eqnarray}
f_1(s,t)&=&a^{-1}-k_{\rm F}[R_\uparrow(s,t)+R_\downarrow(s,t)],\nonumber\\
f_2(s,t)&=&k_{\rm F}I(s,t),
\end{eqnarray}
where $R_\sigma(s,t)$ and $I(s,t)$ are given by
\begin{eqnarray}
R_\sigma(s,t)&=&\frac{\eta_\sigma}{\pi}+\frac{\eta_\sigma^2-(s+t)^2}{4\pi s}\ln\left|\frac{\eta_\sigma+s+t}{\eta_\sigma-s-t}\right|\nonumber\\
&+&\frac{\eta_\sigma^2-(s-t)^2}{4\pi s}\ln\left|\frac{\eta_\sigma+s-t}{\eta_\sigma-s+t}\right|,\nonumber\\
I(s,t)&=&\Theta(\eta_\uparrow^2+\eta_\downarrow^2-2s^2-2t^2)\prod_{\sigma=\uparrow,\downarrow}\Theta(\eta_\sigma-|s-t|)\nonumber\\
&\times&\left[t+\sum_{\sigma=\uparrow,\downarrow}\frac{\eta_\sigma^2-(s+t)^2}{4s}\Theta(s+t-\eta_\sigma)\right].
\end{eqnarray}

Let us focus on the balanced case $x=0$. The functions $f_1$ and $f_2$ at ${\bf P}=0$ ($s=0$) can be simplified as
\begin{equation}
\frac{f_1}{k_{\rm F}}=\frac{1}{g}-\frac{4}{\pi}\left(1-\frac{t}{2}\ln\frac{1+t}{1-t}\right), \ \ \ \ \ \ \frac{f_2}{k_{\rm F}}=t
\end{equation}
for $0<t<1$. Here $g=k_{\rm F}a$ is the gas parameter in 3D. In the weak coupling limit where $g\rightarrow 0$, we have
$\phi_{\rm m}\simeq-\arctan(ka)$, which coincides with the two-body scattering phase shift $\phi_{2 \rm B}(k)$ in the absence of bound state.
However, when approaching the resonance, the nature of the binary interaction is qualitatively changed by the medium effect. At the BEC side ($g>0$) of the resonance, a simple mathematical exercise shows that the function $f_1$ has a zero $t=t_0\in (0,1)$ for $g>\pi/4$. At the BCS side ($g<0$) of the resonance, this zero always exists. In the BCS limit, the zero can be expressed as
\begin{equation}
t_0=\sqrt{1-\frac{\varepsilon_c}{2E_{\rm F}}},
\end{equation}
where $\varepsilon_c\simeq8E_{\rm F}\exp{(\frac{\pi}{2g}-2)}$ is the Cooper-pair binding energy. Here $E_{\rm F}= k_{\rm F}^2/2$ is the Fermi energy of the noninteracting system.

The numerical results for $\phi_{\rm m}$ at ${\bf P}=0$ are shown in Fig. \ref{fig1}(a). Once $f_1$ has a zero $t=t_0$, it is easy to show that $f_1<0$ for $0<t<t_0$ and $f_1>0$ for $t_0<t<1$. Accordingly, we have $\phi_{\rm m}>0$ for $0<t<t_0$ and $\phi_{\rm m}<0$ for $t_0<t<1$. A jump of $\pi$ at $t=t_0$ appears. Therefore, the binary interaction becomes mixed in the region $-\infty<1/g<4/\pi$. It is attractive at low energy ($0<t<t_0$) and repulsive at high energy ($t_0<t<1$). Even though we have shown the results only for ${\bf P}=0$, the qualitative behavior for
${\bf P}\neq0$ is similar. From the results shown in Fig. \ref{fig1}(a), it is intuitive that when approaching the resonance from the BEC side,
the attractive region with $\phi_{\rm m}>0$ becomes larger and larger.

\subsection {Two dimensions}

In 2D, the functions $f_1(s,t)$ and $f_2(s,t)$ become dimensionless. In the phase space ${\cal S}$, they can be expressed as
\begin{eqnarray}
f_1(s,t)&=&-2\ln(ka_2)-[R_\uparrow(s,t)+R_\downarrow(s,t)],\nonumber\\
f_2(s,t)&=&I(s,t),
\end{eqnarray}
where $R_\sigma(s,t)$ and $I(s,t)$ are given by
\begin{eqnarray}
R_\sigma(s,t)&=&\int_0^\pi \frac{d\theta}{\pi}\Theta(\eta_\sigma-s\sin\theta)
\ln\left|\frac{(u_\sigma^+)^2\Theta(u_\sigma^+)-t^2}{(u_\sigma^-)^2\Theta(u_\sigma^-)-t^2}\right|,\nonumber\\
I(s,t)&=&\Theta(1-s^2-t^2)\prod_{\sigma=\uparrow,\downarrow}\Theta(\eta_\sigma-|s-t|)\nonumber\\
&\times&\left[\pi-\sum_{\sigma=\uparrow,\downarrow}\arccos\frac{\eta_\sigma^2-s^2-t^2}{2st}\Theta(s+t-\eta_\sigma)\right].
\end{eqnarray}
Here $u_\sigma^\pm=s\cos\theta\pm(\eta_\sigma^2-s^2\sin^2\theta)^{1/2}$. Note that we have
$-2\ln(ka_2)=2/g-2\ln t$ by using the gas parameter $g=-1/\ln(k_{\rm F}a_2)$ in 2D.

For the balanced case $x=0$, we have
\begin{eqnarray}
f_1=\frac{2}{g}+2\ln\frac{t}{1-t^2},\ \ \ \ f_2=\pi
\end{eqnarray}
in the phase space $0<t<1$ for ${\bf P}=0$. In contrast to the 3D case, the function $f_1$ has a zero $t=t_0\in (0,1)$ for arbitrary values of $a_2>0$. We have $\phi_{\rm m}>0$ for $0<t<t_0$ and $\phi_{\rm m}<0$ for $t_0<t<1$. In the BCS limit $a_2\rightarrow+\infty$, we have $t_0=\sqrt{1-\varepsilon_c/(2E_{\rm F})}$ where $\varepsilon_c\simeq\sqrt{2\varepsilon_{\rm B}E_{\rm F}}$ is the Cooper-pair binding energy in 2D \cite{BCS2D,BCS2DR,MC2D}. Therefore, there exists a qualitative difference between 2D and 3D: the in-medium binary interaction shows attraction at low energy for arbitrary value of $a_2$ in 2D. The results for $\phi_{\rm m}$ are shown in Fig. \ref{fig1}(b). For arbitrary value of the gas parameter $g$, we find that the phase shift $\phi_{\rm m}$ is positive in the region $0<t<t_0$.

\section {Interaction energy}\label{s4}

The interaction energy ${\cal E}_{\rm int}$ can be obtained by summing the perturbative contributions of the GF type to all orders in $g$, which has been done by Kaiser \cite{LRS02} in 3D. Consider the open ladder diagram with $n$ contact interactions. It is roughly given by the ($n-1$)th power of $\Pi$ times $U^n$. When closing the two open fermion lines, we replace them by two medium insertions. This introduces an integration
\begin{eqnarray}
&-&\int_{-\infty}^\infty\frac{dP_0}{2\pi}\int_{-\infty}^\infty\frac{dk_0}{2\pi}\sum_{\bf P}\sum_{\bf k}{\cal G}_{\rm m}^\uparrow\left(k_+,{\bf k}_+\right)
{\cal G}_{\rm m}^\downarrow\left(k_-,{\bf k}_-\right)\cdots\nonumber\\
&=&\sum_{\bf P}\sum_{\bf k}n_\uparrow({\bf k}_+)
n_\downarrow({\bf k}_-)\int_{-\infty}^\infty dP_0\delta\left(P_0-\frac{{\bf P}^2}{4}-{\bf k}^2\right)\cdots.
\end{eqnarray}
The delta function $\delta(P_0-{\bf P}^2/4-{\bf k}^2)$ clearly shows that the interaction energy contains only the contribution from the scattering states.

On the scattering mass shell $P_0={\bf P}^2/4+{\bf k}^2$, we have $\Pi(s,t)=U^{-1}-(f_1+if_2)/(4\pi)$. We note that only the closed ladders that have at least one pair of adjacent MIs contribute to the interaction energy. By using the special property $\Pi-\Pi_2=\Pi^*$ we can take into account the symmetry factors which will also correct for the overcounting of certain diagrams. After a careful combinatorial analysis, we find that the interaction energy density is given by \cite{SUPP}
\begin{eqnarray}
{\cal E}_{\rm int}&=&\sum_{\bf P}\sum_{\bf k}n_\uparrow({\bf k}_+)n_\downarrow({\bf k}_-)
\sum_{n=1}^\infty{\cal C}_n(s,t),
\end{eqnarray}
where the $n$th-order contribution is
\begin{eqnarray}
{\cal C}_n(s,t)=-\frac{[U\Pi(s,t)]^n-[U\Pi^*(s,t)]^n}{2i n}\frac{4\pi}{f_2(s,t)}.
\end{eqnarray}
Completing the summation over $n$, we obtain
\begin{eqnarray}
\sum_{n=1}^\infty{\cal C}_n(s,t)&=&\frac{4\pi}{f_2(s,t)}\frac{1}{2i}\ln\frac{f_1(s,t)+if_2(s,t)}{f_1(s,t)-if_2(s,t)}\nonumber\\
&=&-\frac{4\pi\phi_{\rm m}(s,t)}{f_2(s,t)}.
\end{eqnarray}
The above result shows that the interaction energy is an integration of the phase shift $\phi_{\rm m}$ over the phase space ${\cal S}$.

For the 3D case, the perturbative expansion of the interaction energy can be expressed as
\begin{eqnarray}
{\cal E}_{\rm int}=\sum_{n=1}^\infty c_n(k_{\rm F}a)^n,
\end{eqnarray}
where the expansion coefficients
\begin{equation}
c_n=4\pi\sum_{\bf P}\sum_{\bf k}n_\uparrow({\bf k}_+)n_\downarrow({\bf k}_-)H_n(I,R).
\end{equation}
Here $R(s,t)=R_\uparrow(s,t)+R_\downarrow(s,t)$ and
\begin{equation}
H_n(I,R)=\frac{(R+iI)^n-(R-iI)^n}{2i nI}.
\end{equation}
Up to the sixth order, we have
\begin{eqnarray}
&&H_1=1,\ \ \ \ H_2=R,\ \ \ \ H_3=R^2-\frac{1}{3}I^2,\nonumber\\
&&H_4=R^3-RI^2,\ \ \ \ H_5=R^4-2R^2I^2+\frac{1}{5}I^4,\nonumber\\
&&H_6=R^5-\frac{10}{3}R^3I^2+RI^4.
\end{eqnarray}
By using a detailed diagrammatic analysis, Kaiser has carefully checked up to the sixth order that the above perturbative expansion includes precisely the perturbative contributions from particle-particle ladders, hole-hole ladders, and mixed particle-particle and hole-hole ladders \cite{LRS02}.

\subsection {Three dimensions}

In 3D, the interaction energy density can be expressed as
\begin{eqnarray}
\frac{{\cal E}_{\rm int}}{{\cal E}_0}=-\frac{80}{\pi}\int\int_{\cal S}s^2t \phi_{\rm m}(s,t)dsdt,
\end{eqnarray}
where ${\cal E}_0=\frac{3}{5}nE_{\rm F}$ is the energy density of a noninteracting Fermi gas. For small $g$, we have
\begin{equation}
-\phi_{\rm m}=gI+g^2I(R_\uparrow+R_\downarrow)+O(g^3).
\end{equation}
Using this expansion, we can recover precisely the second-order perturbation theory \cite{LHY,EFT,Kanno}. For the balanced case $x=0$, we obtain
\begin{eqnarray}
\frac{{\cal E}_{\rm int}}{{\cal E}_0}=\frac{10}{9\pi}g+\frac{4(11-\ln 2)}{21\pi^2}g^2+O(g^3).
\end{eqnarray}
For the imbalanced case $x\neq0$, we have
\begin{eqnarray}
\frac{{\cal E}_{\rm int}}{{\cal E}_0}=\frac{10(1-x^2)}{9\pi}g
+\frac{\xi(\eta_\uparrow,\eta_\downarrow)}{21\pi^2}g^2+O(g^3),
\label{second}
\end{eqnarray}
where the second-order coefficient $\xi(\eta_\uparrow,\eta_\downarrow)$ reads
\begin{eqnarray}
\xi&=&22\eta_\uparrow^3\eta_\downarrow^3(\eta_\uparrow+\eta_\downarrow)-4\eta_\uparrow^7\rm{ln}\frac{\eta_\uparrow+\eta_\downarrow}{\eta_\uparrow}-4\eta_\downarrow^7\rm{ln}\frac{\eta_\uparrow+\eta_\downarrow}{\eta_\downarrow}\nonumber\\
&+&\frac{1}{2}(\eta_\uparrow-\eta_\downarrow)^2\eta_\uparrow\eta_\downarrow(\eta_\uparrow+\eta_\downarrow)[15(\eta_\uparrow^2+\eta_\downarrow^2)+11\eta_\uparrow\eta_\downarrow]\nonumber\\
&+&\frac{7}{4}(\eta_\uparrow-\eta_\downarrow)^4(\eta_\uparrow+\eta_\downarrow)(\eta_\uparrow^2+\eta_\downarrow^2+3\eta_\uparrow\eta_\downarrow)\rm{ln}\bigg|\frac{\eta_\uparrow-\eta_\downarrow}{\eta_\uparrow+\eta_\downarrow}\bigg|.
\end{eqnarray}
The perturbative result for $x\neq0$ agrees with the result first evaluated by Kanno \cite{Kanno}.

\begin{figure}[!htb]
\begin{center}
\includegraphics[width=9.3cm]{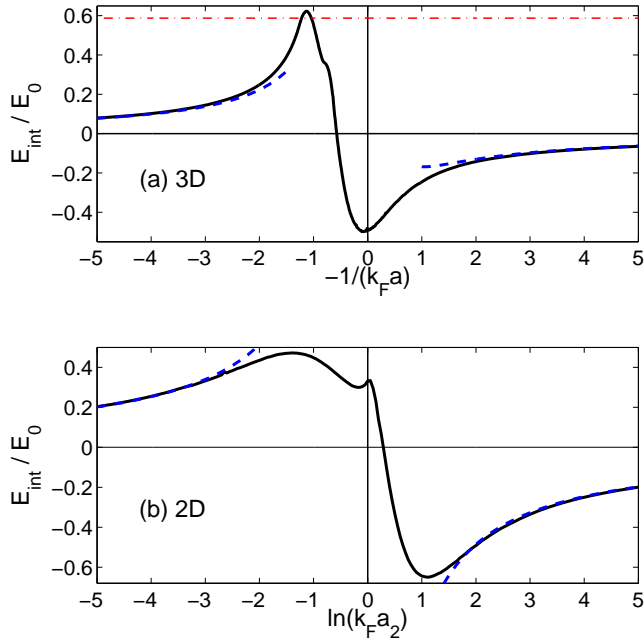}
\caption{(Color online) The interaction energy for the balanced case $x=0$ as a function of $-1/(k_{\rm F}a)$ in 3D (a)
and of $\ln(k_{\rm F}a_2)$ in 2D. The blue dashed lines are the results from second-order perturbation theory. The red dash-dotted
line in (a) corresponds to the energy of the fully polarized state in 3D, ${\cal E}_{\rm fp}=2^{2/3}{\cal E}_0$.  \label{fig2}}
\end{center}
\end{figure}

The interaction energy for the balanced case is shown in Fig. \ref{fig2}(a). It reaches a maximum $0.62{\cal E}_0$ at $g=0.88$ and then decreases. This can be clearly understood by the qualitative change of the binary interaction in medium: the phase shift $\phi_{\rm m}$ experiences more and more attraction when approaching the resonance from the repulsive side $a>0$. We notice that a recent quantum Monte Carlo study of the 3D dilute Hubbard model also found that the interaction energy shows a maximum at some interaction strength \cite{QMCH}. The energy maximum is only $0.034{\cal E}_0$ larger than the energy of the fully polarized state. At unitarity the Bertsch parameter (for the normal phase) reads $\xi=0.507$, which agrees with the experimental result $\xi=0.51(2)$ \cite{KSI01} and the Monte Carlo results: $\xi\simeq 0.54$  \cite{KSI02}, $\xi=0.56$ \cite{KSI03}, and $\xi=0.52$ \cite{KSI04}. Notice that our approach at $T=0$ does not predict any discontinuity of the energy and its slope, in contrast to the results from a generalized NSR approach \cite{Ho}. In particular, we have checked the compressibility $\kappa$ which is defined as
$1/(n^2\kappa)=\partial^2{\cal E}/\partial n^2$. We find that it is always positive in our approach.

It is also intuitively easy to understand the behavior of the interaction energy by using an ``effective" scattering length $a_{\rm eff}$ in the medium. At vanishing center-of-mass momentum ${\bf P}=0$ and at small scattering energy $E={\bf k}^2\ll E_{\rm F}$, the in-medium scattering amplitude can be expressed as
\begin{eqnarray}
f_{\rm m}(k)=\frac{4\pi}{\frac{1}{a}-\frac{4k_{\rm F}}{\pi}+\frac{4}{\pi k_{\rm F}}k^2+ik}.
\end{eqnarray}
In analogy to the vacuum case, the effective scattering length $a_{\rm eff}$ in the medium can be defined as $a_{\rm eff}=1/(a^{-1}-4k_{\rm F}/\pi)$. It is also interesting that the medium effect generates an effective range. The effective scattering length $a_{\rm eff}$ diverges at $g=\pi/4\simeq0.79$. The location of the energy maximum corresponds to $1/(k_{\rm F}a_{\rm eff})=-0.14$, which lies at the ``BCS side" in terms of $a_{\rm eff}$.

\subsection {Two dimensions}

In 2D, the interaction energy density is given by
\begin{eqnarray}
\frac{{\cal E}_{\rm int}}{{\cal E}_0}=-\frac{32}{\pi}\int\int_{\cal S} st \phi_{\rm m}(s,t)dsdt,
\end{eqnarray}
where ${\cal E}_0=\frac{1}{2}nE_{\rm F}$ is the energy density of a noninteracting 2D Fermi gas. For small $g$, we have
\begin{equation}
-\phi_{\rm m}=\frac{1}{2}gI+\frac{1}{4}g^2I(2\ln t+R_\uparrow+R_\downarrow)+O(g^3).
\end{equation}
For the balanced case $x=0$, we obtain
\begin{eqnarray}
\frac{{\cal E}_{\rm int}}{{\cal E}_0}=g+\frac{3-4\ln 2}{4}g^2+O(g^3).
\end{eqnarray}
The coefficient of the second-order term, $(3-4\ln2)/4\simeq0.057$, agrees with the result by Engelbrecht, Randeria, and Zhang \cite{PEOS2D1}
but disagrees with Bloom's numerical result $0.28$ \cite{PEOS2D2}. The 2D scattering length is also defined as $\varepsilon_{\rm B}=4/(a_2^2e^{2\gamma})$ in some papers, where $\gamma\simeq0.577$ is Euler's constant. For this definition, we have
\begin{eqnarray}
\frac{{\cal E}_{\rm int}}{{\cal E}_0}=g+\left(\gamma+\frac{3}{4}-2\ln 2\right)g^2+O(g^3).
\end{eqnarray}
where the second-order coefficient $\gamma+\frac{3}{4}-2\ln 2\simeq-0.059$ becomes negative. For the imbalanced case, we do not have an analytical result for the perturbative expansion.

The interaction energy for the balanced case is shown in Fig. \ref{fig2}(b). It reaches a maximum $0.47{\cal E}_0$
at $g=0.71$ or $\ln(k_{\rm F}a_2)=-1.4$. The energy curve around the maximum becomes much flatter than in the 3D case.
As a result, the energy maximum becomes much lower than the energy of the fully polarized state ${\cal E}_{\rm fp}=2{\cal E}_0$.
These results can be understood intuitively through the behavior of $\phi_{\rm m}$: In 2D, the binary interaction is qualitatively
changed even in the BEC limit $a_2\rightarrow 0^+$.

\section {Itinerant ferromagnetism}\label{s5}

Finally we study the possibility of itinerant ferromagnetism in the many-body scattering state. It is intuitively clear that existence of an energy maximum at the BEC side of the resonance has a significant effect on the itinerant ferromagnetism. To study the itinerant ferromagnetism, we study the system with a finite polarization $x=(n_\uparrow-n_\downarrow)/(n_\uparrow+n_\downarrow)$ and analyze the landscape of the energy density ${\cal E}(x)$.

\subsection {Three dimensions}

\begin{figure}[!htb]
\begin{center}
\includegraphics[width=9.3cm]{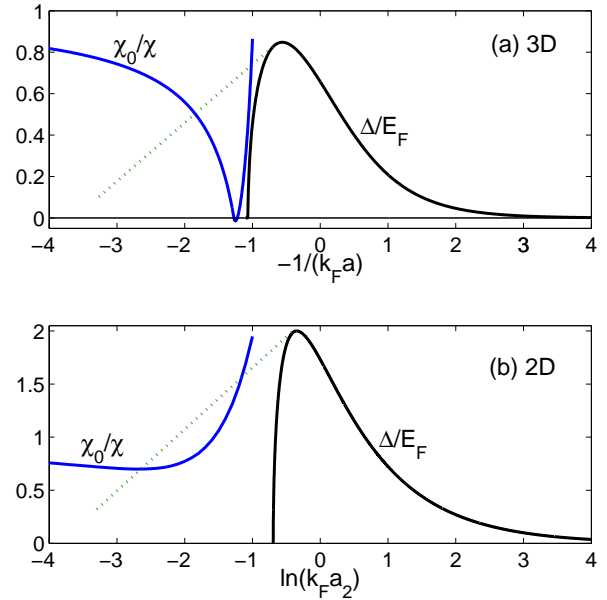}
\caption{(Color online) The normalized inverse spin susceptibility $\chi_0/\chi$ (blue solid lines) and the pairing decay rate $\Delta$  divided by
$E_{\rm F}$ (black solid lines) as functions of the gas parameters in 3D (a) and 2D (b). The green dotted lines show schematically the behavior of
the decay rate when three-body processes are taken into account. \label{fig3}}
\end{center}
\end{figure}

Let us first assume that the many-body scattering state can be prepared in equilibrium. By analyzing the energy curve ${\cal E}(x)$, we find that the system undergoes a second-order phase transition to the ferromagnetic phase at $g=0.79$ where the spin susceptibility $\chi$ diverges and then a first-order order phase transition to the paramagnetic phase at $g=0.96$. This reentrant phenomenon can be clearly understood from the existence of an energy maximum at $g=0.88$. The spin susceptibility $\chi$ can be obtained by making use of a small-polarization expansion of the energy density,
\begin{equation}
{\cal E}(x)={\cal E}(0)+\alpha x^2+\cdots.
\end{equation}
We have $\chi_0/\chi\propto\alpha$. Here $\chi_0$ is the spin susceptibility of noninteracting Fermi gases. The normalized inverse spin susceptibility $\chi_0/\chi$ is shown in Fig. \ref{fig3}(a). In a narrow region $0.79<g<0.82$ where $\chi_0/\chi<0$, the system phase separates into partially polarized domains. We notice that the second-order ferromagnetic transition at $g=0.79$ is very close to the gas parameter $g=\pi/4$ where the in-medium scattering length $a_{\rm eff}$ diverges.

The maximum critical temperature of ferromagnetism $T_c^{\rm max}$ becomes constrained by the energy maximum. Since $\chi_0/\chi>0$ near the energy maximum, $T_c^{\rm max}$ can be roughly estimated by using second-order perturbation theory. By equating the energy of the second-order perturbation theory to the energy maximum ${\cal E}_{\rm max}=1.62{\cal E}_0$, we estimate $T_c^{\rm max}\simeq0.2T_{\rm F}$. Above this temperature, the ferromagnetic phase disappears and one can never observe a diverging spin susceptibility.  We note that
the lowest temperature realized in the first experiment of Ketterle and co-workers \cite{exp} is about $T=0.12T_{\rm F}$ and a later experiment \cite{exp4} at $T=0.23T_{\rm F}$ did not observe any diverging behavior of the spin fluctuation.

On the other hand, the many-body scattering state is not stable and suffers from various decay processes. In the deep BEC region where $g$ is small, it has been shown that the three-body recombination rate is proportional to $\bar{\varepsilon}(na^3)^2$ \cite{Threebody}, where $\bar{\varepsilon}$ is the average kinetic energy of a fermion. In a degenerate Fermi gas at zero temperature, $\bar{\varepsilon}$ is given by $3E_{\rm F}/5$. This indicates that the decay rate of the repulsive branch is quite small for a small positive gas parameter $g$. Recent experiments on the repulsive branch \cite{exp4,exp5} found that equilibrium study of the repulsive Fermi gas is possible only for $g<0.25$ for temperature around $0.3T_{\rm F}$. At large gas parameter $g$, fast decay of the gas prevents the observation of the equilibrium profiles.

For large gas parameter $g$, it seems impossible to present an accurate theoretical study of the decay rate. However, the present many-body approach allows us to study the pair (molecule) formation rate or pairing decay rate from an in-medium two-body picture \cite{Pekker,Macdonald}. It has been shown that this pairing decay picture can qualitatively explain the experimental observations of the fast decay at large $g$ \cite{Pekker}. The pair formation rate is characterized by the imaginary part of the pole of the in-medium $T$ matrix $T_{\rm m}(P_0,{\bf P})$. For a fixed pair momentum
${\bf P}$, we make an analytical continuation of the variable $P_0$ to the complex plane. The pole can be expressed as
$P_0=\Omega_{\bf P}+i\Delta_{\bf P}$, where the imaginary part $\Delta_{\bf P}$ characterizes the pair formation rate \cite{Pekker,Macdonald}.
The strongest decay occurs at ${\bf P}=0$ for balanced populations. The result of the pairing decay rate $\Delta\equiv\Delta_{{\bf P}=0}$ at zero temperature is shown in Fig. \ref{fig3}(a). It arises at $g=0.93$ and rapidly reaches a maximum at $g=1.8$. In the BCS limit, the pairing decay rate coincides with the superfluid gap, $\Delta\simeq 8E_{\rm F}\exp{(\frac{\pi}{2g}-2)}$. The sharp onset at $g=0.93$ is expected to be smoothed by three-body processes, since the three-body decay rate is proportional to $\bar{\varepsilon}(na^3)^2$ for small positive $g$ \cite{Threebody}.

The study of the equilibrium properties of the repulsive Fermi gas is therefore limited within the time scale of pair formation. From the result of the pairing rate $\Delta$ shown in Fig. \ref{fig3}(a), we find that pair formation occurs in a time scale $2\hbar/E_{\rm F}$ for a wide range of the gas parameter around the ferromagnetic phase. For typical atom density $n$ realized in experiments, we estimate that this time scale is of order 0.1 ms, which is every short for experimental observation of the equilibrium profiles. Actually, a recent experiment has observed a rapid decay into bound pairs over times on the order of $10\hbar/E_{\rm F}$ for a wide range of the interaction strength \cite{exp4}. Future experimental studies of repulsive Fermi gases should overcome the fast decay rate. Theoretical studies have suggested several ways to suppress the decay rate: narrow resonance \cite{Pekker2}, high temperature \cite{Pekker}, low dimensionality \cite{Demler}, population imbalance \cite{Macdonald}, mass imbalance \cite{Threebody,Conduit}, and lattice and band structure \cite{Troyer}. It will be interesting to extend the present nonperturbative approach to study the above effects.

\subsection {Two dimensions}

The mean-field theory in 2D predicts a ferromagnetic phase transition at $g=1$ or $\ln(k_{\rm F}a_2)=-1$ since the energy density is given by
\begin{equation}
\frac{{\cal E}_{\rm mf}(x)}{{\cal E}_0}=1+x^2+(1-x^2)g.
\end{equation}
However, the present nonperturbative analysis rather suggests that there exists no itinerant ferromagnetism in a 2D Fermi gas at zero temperature. 
We have carefully studied the energy curve ${\cal E}(x)$ and found that the minimum is always located at $x=0$. The normalized inverse spin susceptibility $\chi_0/\chi$ is shown in Fig. \ref{fig3}(b). It never reaches zero, which indicates no ferromagnetic transition. This can be intuitively understood by the fact that the energy maximum is much lower than the energy of the fully polarized state. We notice that a recent quantum Monte Carlo study of
a two-component Fermi gas with hard-core interactions also suggested an absence of itinerant ferromagnetism in 2D \cite{QMC2D}. The pairing decay rate in 2D can be analytically evaluated as
\begin{equation}
\Delta=\Theta(8E_{\rm F}-\varepsilon_{\rm B})\sqrt{2\varepsilon_{\rm B}E_{\rm F}-\frac{1}{4}\varepsilon_{\rm B}^2},
\end{equation}
which shows a maximum at $\ln(k_{\rm F}a_2)=-0.35$.

\section {Summary}\label{s6}
In this work, we have studied the behavior of the interaction energy and the possibility of itinerant ferromagnetism in a strongly interacting Fermi gas at zero temperature in the absence of molecule formation. The interaction energy of the system is obtained by summing the perturbative contributions of the GF type to all orders in the gas parameter. We show that in both 3D and 2D, the interaction energy arrives at a maximum before reaching the resonance from the BEC side. This phenomenon can be understood qualitatively through the nature of the binary interaction in the medium: the in-medium scattering phase shift shows attraction at low energy and hence reduces the interaction energy before reaching the resonance. The appearance of an energy maximum has significant effects on the possibility of itinerant ferromagnetism in the system we study. In 3D, the ferromagnetic transition is reentrant and itinerant ferromagnetism exists in a narrow range of the interaction strength. In 2D, however, the present nonperturbative many-body approach suggests that itinerant ferromagnetism does not exist, which reflects the fact that the energy maximum becomes much lower than the energy of the fully polarized state.

In this work we have focused on the balanced case $x=0$. It will be interesting to apply the nonperturbative approach to study the highly polarized case ($x\rightarrow1$) and the properties of the polaron \cite{Bruun}.

{\bf Acknowledgments:} We thank Joseph Carlson and Stefano Gandolfi for useful discussions and G. J. Conduit and Georg M. Bruun for helpful communications. The work is supported by the U. S. Department of Energy Nuclear Physics Office, by the topical collaborations on Neutrinos and Nucleosynthesis, and by Los Alamos National Laboratory.

\end{document}